\documentclass{midl} % Include author names
\usepackage{mwe} % to get dummy images

\jmlrproceedings{MIDL}{ARXIV}
\jmlrpages{2}
\jmlryear{2022}

\title[Robust Segmentation]{Learning from imperfect training data using a robust loss function: application to brain image segmentation}

\midlauthor{\Name{Haleh Akrami\midljointauthortext{Contributed equally}\nametag{$^{1}$}} \Email{akrami@usc.edu}\\
\Name{Wenhui Cui\midlotherjointauthor\nametag{$^{1}$}} \Email{wenhuicu@usc.edu}\\
\Name{Anand A. Joshi\nametag{$^{1}$}} \Email{ajoshi@usc.edu}\\
\Name{Richard M. Leahy\nametag{$^{1}$}} \Email{leahy@sipi.usc.edu}\\
\addr $^{1}$ University of Southern California, Los Angeles, CA, USA \\
}

\begin{document}

\maketitle

\begin{abstract}
Segmentation is one of the most important tasks in MRI medical image analysis and is often the first and the most critical step in many clinical applications. In brain MRI analysis, head segmentation is commonly used for measuring and visualizing the brain's anatomical structures and is also a necessary step for other applications such as current-source reconstruction in electroencephalography and magnetoencephalography (EEG/MEG). Here we propose a deep learning framework that can segment brain, skull, and extra-cranial tissue using only T1-weighted MRI as input. In addition, we describe a robust method for training the model in the presence of noisy labels.
\end{abstract}

\begin{keywords}
brain, segmentation, robustness, machine-learning
\end{keywords}

\section{Introduction}

Image segmentation is an important task in medical image processing.   Deep learning methods can be used for this purpose but in practice, especially in large datasets, training data inevitably contain mislabeled data, anomalies, or outliers. Training based on noisy labeled datasets can result in output quality loss because deep neural networks (DNNs) can overfit to noisy labels. This issue is particularly severe in the field of medical image analysis where the quality of annotation is strongly dependent on the skill, experience, and time expended by the annotators.

Here our goal is MRI head segmentation when only a T1-weighted image is available. Head segmentation (including gray and white matter, CSF, skull, scalp, and other extra-cranial tissue) is commonly used for measuring and visualizing the brain's anatomical structures and is also a necessary step for other applications, for example in generating head-models for use in current-source reconstruction in electroencephalography and magnetoencephalography (EEG/MEG) \cite{huang2015fully}.

T1-weighted (T1-W) images are commonly available and employ the most cost-effective and fastest pulse sequences used to create accurate head segmentations. However, adding T2-weighted (T2-W) images can help better segmentation of the skull due to superior contrast, particularly between CSF and skull \cite{nielsen2018automatic}. Our goal was to develop a fast head segmentation using only the T1-W modality. We investigated three scenarios: (i) First, we generated head segmentations ("software-generated labels", (SGLs) using SimNIBS (\url{https://simnibs.github.io}) where the input was T1-W and T2-W images. We then trained a CNN using only T1-W as input to predict these labels. (ii) We obtained SGLs using only T1-W data and trained a CNN using T1-W as input to predict these labels. (iii) We repeated (ii) using a robust loss function to compensate for the poor quality of training labels for some classes due to the absence of T2-W images.

%Methods for mitigating the effect of noisy labels can be categorized into three classes: (1) label correction methods that improve the quality of raw labels by modeling characteristics of the noise and correcting incorrect labels \cite{xiao2015learning}; (2) methods with robust loss that are inherently robust to labeling errors \cite{wang2019symmetric}; and (3) refined adaptive training strategies that are more robust to noisy labels \cite{yu2019does}. 

Robust loss functions offer a statistically principled approach to the noisy label problem. By applying a power-law function, beta-cross entropy (BCE) loss can mitigate the effect of noise in the training data \cite{akrami2022robust}. Minimizing BCE is equivalent to minimizing beta-divergence, which is the robust counterpart of KL-divergence. In contrast to this earlier work, which focused on anomaly detection, here we focus on image segmentation.

Our key contributions are as follows: (1) training a state-of-the-art neural network, TransUNet \cite{chen2021transunet}, for head segmentation. This method is very fast at inference time compared to traditional software-generated segmentations (less than 1 min. for the neural network approach vs. 5-10 hrs for traditional software); (2) Investigating using of a robust loss in the absence of good quality training data, such as lack of good quality segmentations for training due to the absence of T2-W modality data.

\section{Method}
We used the CamCan dataset (\url{https://www.cam-can.org }) that consists of T1-W and T2-W MRI images. We used SimNIBS to segment the head into 9 classes: 0=background (outside head), 1=WM, 2=GM, 3=CSF, 4= bone, 5= skin, 6= cavities, 7 = eyes, 8=ventricles. We split the data 60\%/20\%/20\% for training/validation/test sets. We trained a TransUNet \cite{chen2021transunet} using 2D slices as input and resized to 256*256 for 10 epochs.
(i) In the first experiment, we generated SGLs using T1-W and T2-W images as input to SimNIBS to segment the head. We then used T1-W as input and T1-W+T2-W SGLs as the ground-truth labels for training the TransUNet (T1-W+T2-W TransUNet). (ii) We obtained T1-W SGLs by using T1-W images only as input to SimNIBS; then, we used T1-W as input and T1-W SGLs as noisy labels for training the TransUNet with a cross-entropy loss (T1-W TransUNet) (iii) finally we repeated experiment (ii)  but with the robust BCE loss (T1-W TransUNet Robust) . The BCE loss is expressed as:
\begin{equation}
\label{eq:bce}
    \mathcal{L}_{BCE}=\frac{{\beta+1}}{\beta} (1-p(y|x))^\beta+ \sum_{k=1}^{K} p(k|x)^{\beta+1}
\vspace{-5pt}
\end{equation}
where $x$ is the input variable, $y$ is the response variable, and $\beta$ is a hyper-parameter; $p(k|x)$ is the probability output of a \textit{deep neural network}~(DNN) classifier, and $K$ is the number of classes. BCE minimizes $\beta$-divergence between the posterior and empirical distributions when the posterior is a categorical distribution \cite{akrami2022robust}. Using L’Hôpital’s rule, it can be shown that BCE is equivalent
to CE for $\lim \beta \to 0$ where $\beta$-divergence also converges to KL-divergence. $\beta$-divergence is the robust counterpart of KL-divergence using a power-law function. We tuned parameter $\beta=0.0001$ using the validation data. We warmed up the robust model for two epochs using the cross-entropy loss. 
\begin{figure}[!htb]%[htbp]
 % Caption and label go in the first argument and the figure contents
 % go in the second argument
\floatconts
  {fig:lables}
  {\vspace{-20pt}\caption{Segmentation results using different methods}}
  {\includegraphics[width=1\linewidth]{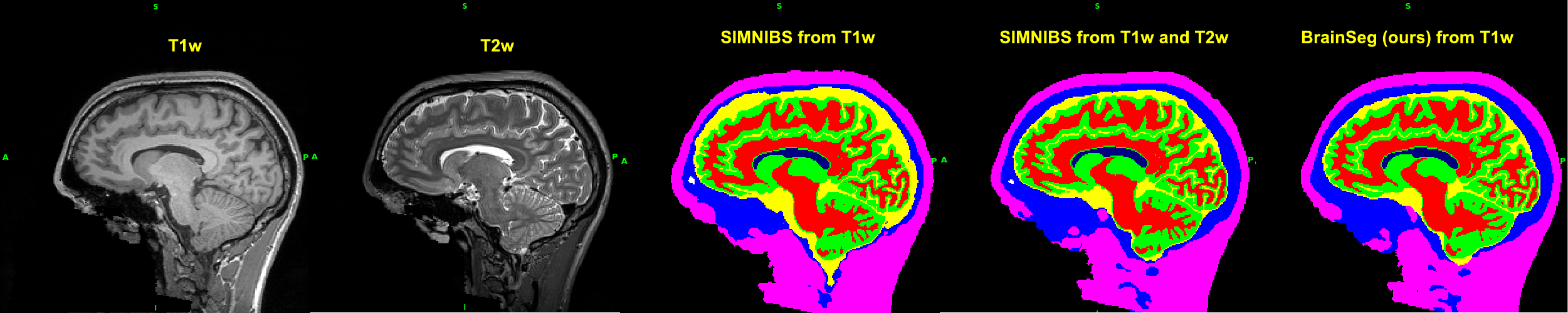}}
  \vspace{-5pt}
  \end{figure}
 \section{Results and discussion}
To measure the performance of our method, we calculated the Dice score between each segmentation and T1-W+T2-W SGL labels (table \ref{tab:dice}). The result shows our T1-W+T2-W TransUNet model can be used to predict head segmentation when only a T1-W image is available during inference. The dice score of bone/skull for this model is improved compared to T1-W SGL segmentation. Further, as we expected we find an improvement in the dice score for the robust model trained with T1-W only SGL labels compared to the non-robust model for these labels. The dice score between T1-W+T2-W SGL and T1-W SGL represents the level of inaccuracy in each label. The degradation of dice coefficients for the eye labels when using the robust loss is probably due to the low occurrence of these voxels in the training data leading to their interpretation as outliers. A hybrid CE/BCE loss would address this issue.
%\vspace{-5pt}
\begin{table}[htbp]
 % The first argument is the label.
 % The caption goes in the second argument, and the table contents
 % go in the third argument.
%\floatconts
 %{tab:F1luad}%
  {\caption{Dice scores relative to ground-truth T1-W+T2-W SGLs for the test dataset}}%
  \label{tab:dice}
  \resizebox{\textwidth}{!}{\begin{tabular}{llllllllll}
  \bfseries Model   & \bfseries background & \bfseries WM  & \bfseries GM   &\bfseries CSF & \bfseries  bones  & \bfseries  skin & \bfseries cavities & \bfseries eyes & \bfseries ventricles  \\ \hline
 T1-W SGL  & 0.99& 0.99 & 0.98 & 0.82 & 0.84 & 0.94 & 0.73 & 0.75 & 0.99 \\ \hline
 T1-W+T2-W TransUnet & 0.99 & 0.92 & 0.90 &  0.81 &  0.89 & 0.94 &
 0.71 & 0.78 & 0.94 \\ \hline
 T1-W TransUnet & 0.99 & 0.93 & 0.90 & 0.74 &  0.83 & 0.93&
 0.68 & 0.74 & 0.94\\ \hline
 T1-W TransUnet Robust& 0.99 & 0.90 & 0.87 &  0.78 & 0.85 &  0.94 & 0.72 & 0.68 & 0.94 \\ \hline
  \end{tabular}}
\end{table}
% Acknowledgments---Will not appear in anonymized version
%\midlacknowledgments{We thank a bunch of people.}
 \vspace{-20pt}
{\bibliography{midl-samplebibliography}}
\end{document}